\documentclass[conference,10pt,letterpaper]{IEEEtran}
\usepackage{cite,amsmath,amssymb,amsfonts,algorithmic,multirow,graphicx,textcomp,xcolor,booktabs,listings,url}
\usepackage[ruled,linesnumbered]{algorithm2e}
\usepackage[font=footnotesize]{caption}
\usepackage[T1]{fontenc}

\begin{document}

\title{ENTS: An Edge-native Task Scheduling System for Collaborative Edge Computing
}
\author{Mingjin~Zhang\textsuperscript{\textsection},~Jiannong~Cao\textsuperscript{\textsection},~Lei~Yang\textsuperscript{\textdagger},~Liang~Zhang\textsuperscript{\textdaggerdbl},~Yuvraj~Sahni\textsuperscript{\textsection},~Shan~Jiang\textsuperscript{\textsection}\\
\textsuperscript{\textsection}Hong Kong Polytechnic University, \textsuperscript{\textdagger}South China University of Technology, \textsuperscript{\textdaggerdbl}Shanghai Jiao Tong University\\ 
\textsuperscript{\textsection}\{csmzhang, csjcao, csysahni, cssjiang\}@comp.polyu.edu.hk, \textsuperscript{\textdagger}sely@scut.edu.cn, \textsuperscript{\textdaggerdbl}zhangliang@sjtu.edu.cn
}
\maketitle
\renewcommand{\thefootnote}{}
\footnotetext{Corresponding author: Shan Jiang}
\thispagestyle{plain}
\pagestyle{plain}

\begin{abstract}
Collaborative edge computing (CEC) is an emerging paradigm enabling sharing of the coupled data, computation, and networking resources among heterogeneous geo-distributed edge nodes. Recently, there has been a trend to orchestrate and schedule containerized application workloads in CEC, while Kubernetes has become the de-facto standard broadly adopted by the industry and academia. However, Kubernetes is not preferable for CEC because its design is not dedicated to edge computing and neglects the unique features of edge nativeness. More specifically, Kubernetes primarily ensures resource provision of workloads while neglecting the performance requirements of edge-native applications, such as throughput and latency. Furthermore, Kubernetes neglects the inner dependencies of edge-native applications and fails to consider data locality and networking resources, leading to inferior performance.
In this work, we design and develop ENTS, the first edge-native task scheduling system, to manage the distributed edge resources and facilitate efficient task scheduling to optimize the performance of edge-native applications. ENTS extends Kubernetes with the unique ability to collaboratively schedule computation and networking resources by comprehensively considering job profile and resource status. We showcase the superior efficacy of ENTS with a case study on data streaming applications. We mathematically formulate a joint task allocation and flow scheduling problem that maximizes the job throughput. We design two novel online scheduling algorithms to optimally decide the task allocation, bandwidth allocation, and flow routing policies. The extensive experiments on a real-world edge video analytics application show that ENTS achieves 43\%-220\% higher average job throughput compared with the state-of-the-art.
\end{abstract}


\section{Introduction}
Recently, there has been a noticeable shift to migrate the computation-intensive workloads from the remote cloud to near-end edges \cite{shi2016edge}. Compared with traditional cloud computing, the emerging edge computing paradigm enjoys outstanding benefits, including reduced response latency and enhanced privacy preservation \cite{mao2017survey}\cite{chen2019deep}. A large number of latency-sensitive and mission-critical applications gradually switch to the deployment at the network edge, e.g., virtual reality \cite{zhang2017towards}, autonomous driving \cite{liu2019edge}, and personalized healthcare \cite{sacco2020edge}. Collaborative edge computing (CEC) is a popular and new edge computing paradigm enabling sharing of data, computation, and networking resources among geo-distributed and heterogeneous edge nodes, including edge servers, edge gateways, and mobile phones \cite{zhang2022eaas}. CEC is promising and beneficial because it provides higher reliability and lower latency and facilitates collaboration among different stakeholders \cite{ning2018green}.

Task scheduling is a fundamental problem of collaborative edge computing, which refers to the arrangement of the user-generated application tasks to the heterogeneous edge nodes by deciding when, where, and how to offload the tasks and how to manage and utilize the underlying computation, storage, and networking resources \cite{mao2017survey}. Many works have investigated the task scheduling problems in collaborative edge computing \cite{meng2019dedas}. Recently, there has been a trend of scheduling containerized application workloads among the geo-distributed and heterogeneous edge infrastructure \cite{zhang2021joint}. This is because the container technology provides lightweight resource virtualization and enables fast application development and flexible service deployment over heterogeneous edge nodes.

There are several solutions to orchestrate containerized applications, such as Swarm \cite{soppelsa2016native}, Kubernetes \cite{luksa2017kubernetes}, and Mesos \cite{hindman2011mesos}. Among them, Kubernetes has established its leadership \cite{burns2016borg}. Many works have studied optimizing the Kubernetes scheduler for the cloud environment, where cloud servers with abundant computation resources are interconnected with a high-bandwidth and stable network in a data center \cite{brewer2015kubernetes}. However, Kubernetes is designed not dedicated to edge computing, neglects the unique features of edge nativeness, and lacks adequate support for edge-native applications \cite{han2021tailored}.

First, edge-native applications are usually performance-aware, demanding high throughput, low latency, and strict privacy. The Kubernetes scheduler is mainly designed to ensure resource provision of workloads, such as the capacity of requested memory and CPU cores. It lacks support to meet the performance requirements of edge-native applications. Second, edge-native applications are with inner dependencies. Many intelligent edge applications are resource-greedy and complex, consisting of lots of inter-dependent components which are usually deployed to multiple edge nodes considering the constraint resource of a single node. However, the Kubernetes scheduler fails to consider the application's inner structure. Third, the data, computation, and networking resources are heterogeneous and coupled with each other. Application deployed on heterogeneous edge nodes experiences distinct performance, and the coupled resources require joint orchestration. However, Kubernetes concentrates on orchestrating computation resources without jointly considering the data locality and networking resources, which may lead to underutilized resources and poor performance of workloads. Though some works \cite{rossi2020geo}\cite{wojciechowski2021netmarks} consider the inner dependencies of workloads and the computation resources among edge nodes for optimizing the application performance, they fail to consider the data locality and resource heterogeneity. 

In this work, we designed and developed ENTS, the first edge-native task scheduling system, to manage the geo-distributed and heterogeneous resources of edge infrastructures and enable efficient task scheduling among distributed edge nodes to optimize application performance. ENTS is developed based on Kubernetes, allowing Kubernetes to collaboratively schedule computation and networking resources considering both job profile and resource status. Specifically, to parse the inner dependencies of the user-submitted jobs, we adopt a data flow programming model, where each task in a job is programmed as a functional module. A profiler is designed to profile the job's execution time on heterogeneous edge nodes. The job profile information will later be used to facilitate efficient task scheduling. We also developed a network manager to manage the networking resources, which collaborates with the Kubernetes original components to jointly orchestrate the coupled resources under the coordination of a newly designed collaborative online scheduler. The scheduler runs the intelligent scheduling algorithms to generate the task scheduling policies to optimize the application performance.

To showcase the efficacy of ENTS, we formulate a joint task allocation and flow scheduling problem for data streaming applications as a case study. The problem is a mixed integrated non-linear problem proven to be NP-hard \cite{liu2020dependency}. We design two online algorithms to solve the problem, which decides how to partition the job, where to allocate the tasks, and how to allocate the routing path and bandwidth for intermediate data flow to optimize the average job throughput. The efficacy of the proposed system is illustrated by developing a real-world testbed for a representative edge video analytics application, namely, object attribute recognition. We develop a real-world hybrid testbed with both physical and virtual edge nodes to evaluate the system even in large scale. Online jobs will continuously arrive and be partitioned and scheduled among the edge nodes. We have comprehensively evaluated the performance of the designed system by comparing it with the state-of-the-art regarding different metrics, including average job throughput and average waiting time. The evaluation results show that our edge-native task scheduling approach improves the performance significantly.

The main contributions of this work are as follows:
\begin{itemize}
  \item We design and develop ENTS to manage the data, computation, and networking resources in the heterogeneous geo-distributed edge infrastructure. ENTS is the first work to jointly manage coupled edge resources for optimizing the performance of edge-native applications.
  \item We formulate a joint task allocation and flow scheduling problem for data streaming applications and propose two online algorithms to solve the problem.
  \item We evaluate the performance of proposed solutions in a real-world testbed with a video analytics application. The experimental results indicate the superiority of ENTS over the baseline approaches in terms of higher job throughput and lower latency.
\end{itemize}


\section{Background and Motivations}\label{sec:back-moti}
In this section, we introduce some background knowledge of the Kubernetes scheduler and illustrate the motivations for designing ENTS through some concise examples.

\begin{figure}[t]
    \centering
    \includegraphics[width=0.83\linewidth]{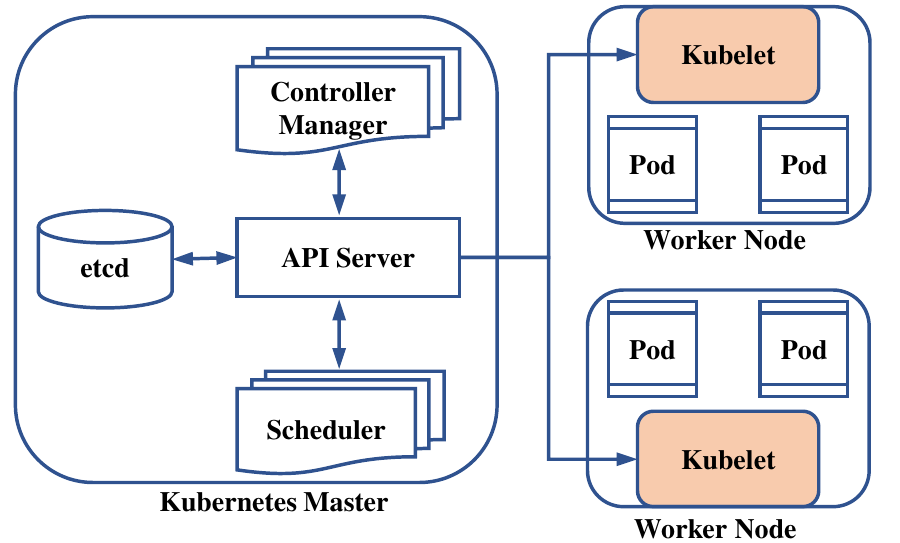} 
    \caption{Components of Kubernetes System}
    \label{k8s_architecture}
\end{figure}

\begin{figure*}[t]
        \includegraphics[width=0.78\linewidth]{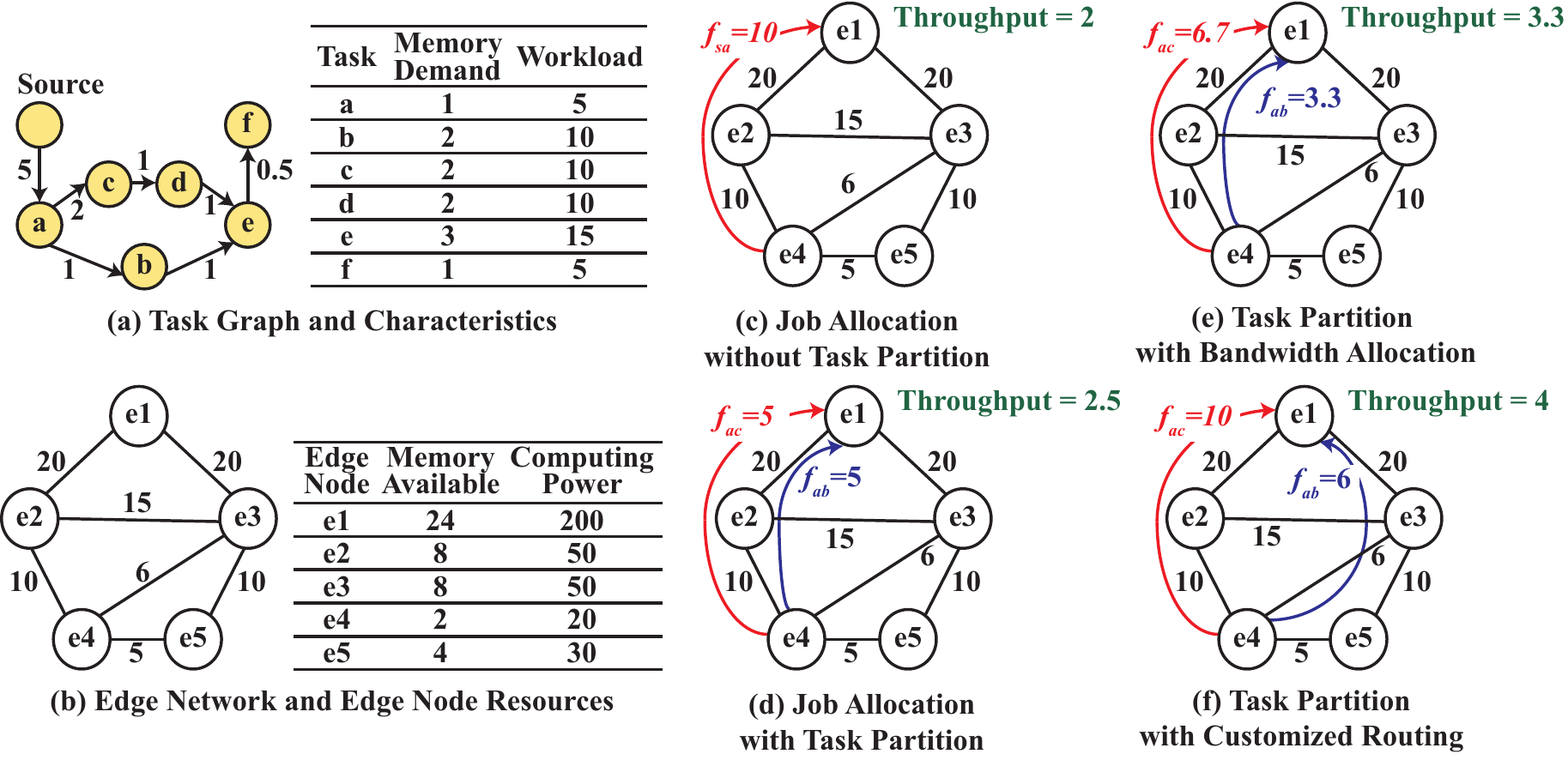} 
        \centering
        \caption{A Motivating Example of Collaborative Task Scheduling}
        \label{motivation}
\end{figure*}

\subsection{Kubernetes Scheduler}
Fig.~\ref{k8s_architecture} depicts the components of Kubernetes with a master-client architecture. There is at least one centralized master managing resources and scheduling containerized workloads across multiple worker nodes (clients). The pod is the basic unit of Kubernetes to schedule the workload. A pod can contain one or more containers. There are mainly four components in the master node. The API server is an entry point to manage the whole cluster, providing services via Restful APIs. Components communicate and interact with each other through the API server. Etcd is a key-value pair distributed database that records the cluster status, such as node resource availability, location, states, and namespace. The scheduler is responsible for scheduling pods. It parses the operational requirements of pods and binds a pod to the best fit node. The controller manager is responsible for monitoring the overall state of the cluster. It launches a daemon running in a continuous loop and is responsible for collecting cluster information. Kubelet is the node agent in the clients. It is responsible for reporting events and resource usage and managing containers. 

When scheduling user-submitted workloads, the scheduler first takes a pod pending to be scheduled from the etcd database and then binds the pod to the corresponding client node according to the pre-defined scheduling policies. The scheduling policy is sent to Kubelet on the client nodes via the API server. After receiving the policies, Kubelet lunches the pods and monitors the pods' execution status. Kubernetes scheduler adopts a multi-criteria decision-making algorithm in two stages. The first stage is node filtering, where the scheduler will select candidate nodes capable of running the pods by applying a set of filters, such as memory and storage availability. Those filters are also known as predicates. The second stage is node scoring. It scores all the candidates based on one or more strategies, such as LeastRequestedPriority, which allocates pods to the nodes with the least computation resource consumption, and BanlancedResourceAllocation, which balances the resource consumption among edge nodes. Those strategies are known as priorities. The scheduler will allocate a pod to the node with the highest score.

\subsection{A Motivating Example}
As shown in Fig.~\ref{motivation}, this section presents a motivating example articulating the benefits of collaborative task scheduling, which jointly considers the coupled data, computation, and networking resources in edge computing scenarios. The problem is to allocate the application with dependent tasks, shown in Fig.~\ref{motivation}(a), to a set of edge nodes, shown in Fig.~\ref{motivation}(b), such that the job throughput is maximized. Fig.~\ref{motivation}(a) shows the task graph of the job modeled as a directed acyclic graph. There are $6$ tasks in the job, and the weight of the link between tasks indicates the volume of the dependent data. Fig.~\ref{motivation}(a) also shows the memory demand and workload of each task. We assume that the total memory demand and workload are the sum of tasks, i.e., $11$ and $55$, respectively. Note that the job is a streaming application, where input data continuously arrives from the source, i.e., edge node $e4$. The amount of the input data is $5$. In Fig.~\ref{motivation}(b), there are $5$ edge nodes $\{e1, e2, e, e4, e5\}$. The weight of the link between the edge nodes indicates the bandwidth. Similarly, the table in Fig.~\ref{motivation}(b) shows the available memory and computing power of the edge nodes in the network. 

Fig.~\ref{motivation}(c) shows the job allocation strategy without task partition, where the job is scheduled to node $e1$ and the input data is transmitted from the source node $e4$ to $e1$ indicated by data flow $f_{sa}$, whose allocated bandwidth is $10$ and routing path is $e4 \rightarrow e2 \rightarrow e1$. The throughput is calculated by $1/\max\{5/10, 55/200\} = 2$. Strategy in Fig.~\ref{motivation}(c) is known as LeastRequestPriority, which are extensively used in Kubernetes. Differently, Fig.~\ref{motivation}(d) partition the job, where task $a$ is allocated to source node $e4$ and the rest tasks are allocated to node $e1$. Hence there are two data flows indicated by $f_{ac}$ and $f_{ab}$ with the same routing path $e4 \rightarrow e2 \rightarrow e1$. By default, two data flows equally share the bandwidth of link $<e2, e4>$. The throughput of the job using this strategy is $2.5$, which is better than strategy in Fig.~\ref{motivation}(c) as the raw data transmission in (c) becomes the bottleneck. Further, Fig.~\ref{motivation}(e) improves (d) with the throughput 3.3 due to the optimized bandwidth sharing policy, where the bandwidths allocated to flow $f_{ac}$ and $f_{ab}$ are proportional to the amount of dependent data. Fig.~\ref{motivation}(f) shows a throughput of 4 with customized routing policy, where the flow $f_{ac}$ selects the routing path $e4 \rightarrow e2 \rightarrow e1$ with the allocated bandwidth $10$ and the flow $f_{ab}$ selects the path $e4 \rightarrow e3 \rightarrow e1$ with the allocated bandwidth $6$. 

From the above examples, we can see that joint consideration of the coupled resources by optimizing the task allocation strategies, the bandwidth allocation, and flow routing policies can improve the application performance. In the rest of this paper, we build ENTS system to orchestrate coupled edge resources and design optimal collaborative task scheduling algorithms by jointly considering the data, computing, and networking resources of the geo-distributed edge nodes.

\section{System Overview}\label{sec:sys-overview}
This section gives an overview of the design goals and the system components. ENTS is designed based on Kubernetes to manage the resources and schedule the workloads over the geo-distributed, large-scale, and heterogeneous edge environment. It has two main objectives: 1) Jointly manage and orchestrate the coupled and distributed data, computation, and networking resources; 2) Enable effective distributed task execution to achieve better performance of applications.

\subsection{Design Goals}
The design of ENTS obeys the principles as follows.
\begin{itemize}
    \item \textit{Scalability}. The system can be scaled to a large number of devices and services retaining its high performance. 
    \item \textit{Collaboration}. The different edge nodes can collaborate to manage the distributed and heterogeneous resource regarding data, computation, and networking.
    \item \textit{Universality}. The system supports execution of various kinds of tasks and workloads.
\end{itemize}

\subsection{System Architecture}
\begin{figure}[t]
        \includegraphics[width=.8\linewidth]{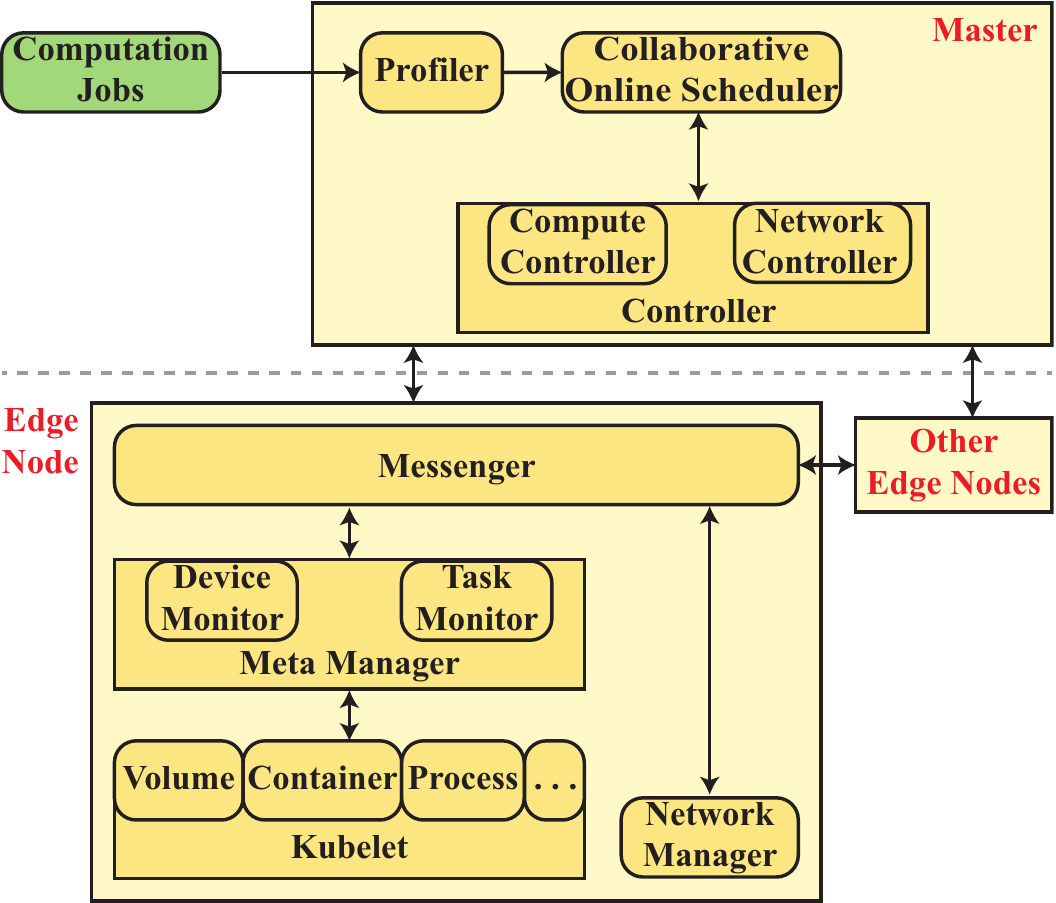} 
        \centering
        \caption{Architecture of the ENTS System}
        \label{architecture}
\end{figure}

In Fig.~\ref{architecture}, we show a birds-eye view of ENTS's system architecture and functional workflow. The system adopts the server-client architecture and is built based on Kubernetes with a master node to manage the distributed resources and schedule the tasks among edge nodes. Kubernetes components are used to manage the computation and storage resources of edge nodes. However, Kubernetes lacks support to profile the job's inner-dependency and execution time on heterogeneous edge nodes and orchestrate networking resources. Hence, we develop new components to enhance the ability of Kubernetes to orchestrate coupled resources considering the job profile. The system follows the principles of service-oriented architecture, where functions of the components are developed as services and can be called with APIs. 

The components of the system are listed below.
\begin{itemize}
    \item \textit{Profiler} parses the input job and profiles the execution time of tasks on heterogeneous edge nodes. The job profile will be used to support intelligent task scheduling.
    \item \textit{Scheduler} accesses the system information, such as CPU and GPU usage, network conditions, and job profile. On this basis, it generates the policies of task execution and resource allocation that optimizes job performance.
    \item \textit{Compute controller} manages the computation and storage resources at the edge nodes. It leverages the Kubernetes components API server and controller manager to orchestrate the computation resources.
    \item \textit{Network controller and manager} manage the networking resources of edge nodes, such as bandwidth allocation, routing and forwarding of data flows.
     \item \textit{Messenger} handles the message between the edge node and the master. We extend the messaging of Kubernetes between the master and clients because it lacks support for orchestrating network resources. 
    \item \textit{Kubelet} manages pods, containers, and data volumes. It is Kubernetes original component, whose primary responsibility is for task execution.
    \item \textit{MetaManager} is responsible for monitoring and storing device status and application status. Specifically, the device and task monitors are responsible for storing and retrieving metadata (device status and task execution status) to and from a lightweight database. Such information will be sent to the master node for supporting task scheduling. 
\end{itemize}

ENTS is based on Kubernetes and reuses the key components of Kubernetes. It equips Kubernetes with the ability to jointly orchestrate the networking and computation resources to optimize the performance of edge-native applications. The general workflow of the system is described as follows. The profiler first parses the user-submitted job and profiles the execution time of each task of the job on heterogeneous edge nodes. The job profile information, including the inter-dependencies of tasks and task execution time, will be used for later decision-making of task scheduling. The scheduler generates the task execution policies by jointly considering the job profile information, the data locality, available computation and networking resources of the edge nodes. Specifically, the policies decide which node to allocate tasks, the bandwidth allocation and the routing path of dataflows. The policies will be managed by the network controller and the compute controller together, and then be executed by the Kubelet and the network manager on the client nodes. The run-time characteristics of tasks and the nodes' status will be sent back to the controller in the master and used for later task scheduling.

\section{System Design}\label{sec:sys-design}
This section illustrates the details of the ENTS system workflow, including job profiling, collaborative task scheduling, and distributed task execution.

\subsection{Application Development and Profiling}
To easily parse the user-submitted job and facilitate efficient distributed task execution, we adopt the data flow programming model \cite{johnston2004advances}, where each task in a job is programmed as a function module. Tasks are loosely coupled with intermediate data transmission. Note that many modern applications are modeled in such a way. Those applications are complex in nature, structured on microservices architecture style, consisting of a large number of inter-dependent and loosely coupled modules. Besides, to support various kinds of workloads, the programming model is non-intrusive to the user programming language. As shown in Fig.~\ref{f: application-declaration}, we only require developers to declare the tasks in the submitted job without intruding on the main functions of the applications. Users can use any programming language to implement their applications. Compared with those programming models, which require users to learn lots of pre-defined operations, such as Hadoop, Spark, and Flink, ENTS is easier to learn and use.   

Users are required to submit the job configuration so that the system can profile the job and perform efficient task scheduling. As shown in Fig.~\ref{f: application}, the configuration explicitly defines the data source, dependencies among the tasks, and the resource demand of each task. Particularly, the job consists of $4$ tasks. The first task $task0$ demands $2$GB memory and has subsequent tasks $task1$ and $task2$. After the user submits the job configuration, ENTS will start the profiling. The objective of job profiling is to estimate the running time of each task of the submitted job on heterogeneous edge nodes, which will then be used to support the collaborative task scheduling. Since it may take much time to profile the job, depending on the complexity of the job, we do the profiling offline. Specifically, the profiler will send the job configuration to the edge nodes that meet the resource requirements of the job. Each edge node will profile the job by executing the tasks under the requested resource and send the job profile information back to the scheduler. Offline profiling is reasonable for those long-running jobs, such as video analytics \cite{zhang2019edge} and virtual reality \cite{zhang2017towards}. Other methods can be used to measure the computing capability of edge nodes and estimate the workload of the application in advance, which is more suitable for online application profiling \cite{kwon2013mantis}\cite{pham2017predicting}. We will study them in the future and incorporate the mechanisms into ENTS.  

\begin{figure}[t]
    \centering
\begin{lstlisting}
# Start definition of tasks
def task_0(input):
  # User code here...
  return True, output

def task_1(input):
  # User code here...
  return True, output

def task_2(input):
  # User code here...
  return True, output

def task_3(input):
  # User code here...
  return True, output

# Export functions as job
job = [task_0, task_1, task_2, task_3]
\end{lstlisting}
    \caption{Code Snippet of User Application}
    \label{f: application-declaration}
\end{figure}

\begin{figure}[t]
    \centering
\begin{lstlisting}
{
  job: "test",
  image: "userid/ents:ubuntu",
  total_memory_request: "4GB",
  source: "",
  input_size: "10"
},
{
  id: "task0",
  downstream: "['task1','task2']",
  memory_resource: "2GB"
},
{
  id: "task1",
  downstream: "['task3']",
  memory_resource: "2GB"
},
{
  id: "task2",
  downstream: "['task3']",
  memory_resource: "2GB"
},
{
  id: "task3",
  downstream: "",
  memory_resource: "2GB"
}
\end{lstlisting}
    \caption{Code Snippet of Application Configuration}
    \label{f: application}
\end{figure}

\begin{figure*}[t]
    \includegraphics[width=.848\linewidth]{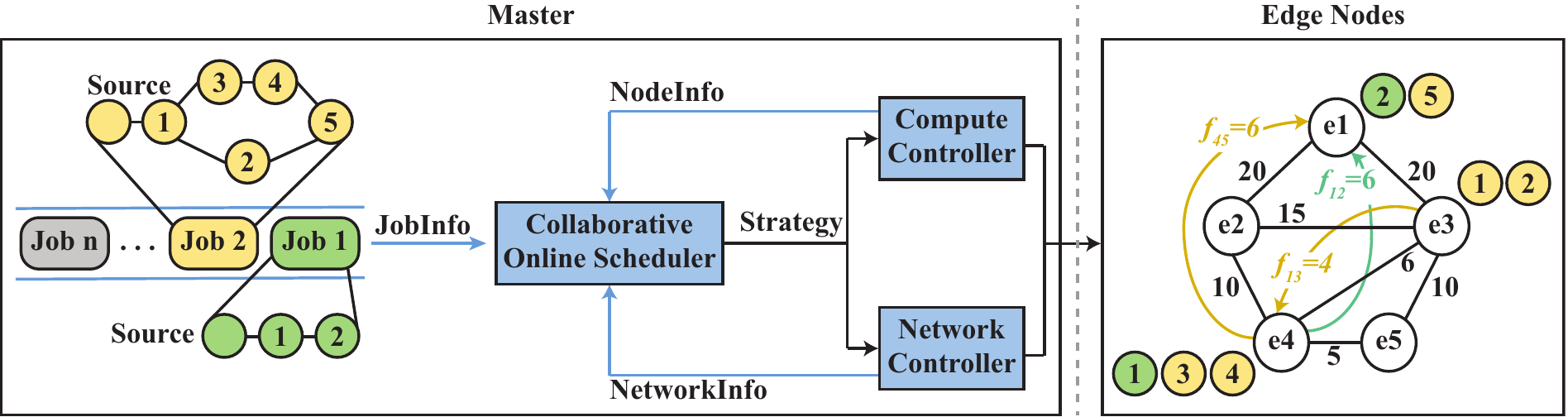} 
    \centering
    \caption{ENTS Task Scheduling Workflow}
    \label{workflow}
\end{figure*}

\subsection{Collaborative Task Scheduling}
After a job is profiled, it will be added to a Job Queue and pending to be scheduled, as shown in Fig.~\ref{workflow}. The job-related information, including task dependencies and requested resources, the available computation resource of edge nodes, and the status of the network will be sent to the scheduler to support the collaborative task scheduling decisions. We will elaborate on the scheduling algorithms in Sec.~\ref{sec:Collaborative-Task-Scheduling}. 

\begin{figure}[t]
    \centering
\begin{lstlisting}[language=Python]
{
  job_name: "test",
  tasks: {
    task_1: {
      task_id: "0", 
      source_node: "edge_1", 
      source_node_port: "8089", 
      previous_node: "", 
      next_node: "edge_2 edge_3", 
      next_node_ports: "8090 8091", 
      bandwidth: "15Mbps 10Mbps", 
      routing: ""},
    task_2: {
      task_id: "1", 
      source_node: "edge_2", 
      source_node_port: "8090", 
      previous_node: "edge_1", 
      next_node: "edge_3", 
      next_node_ports: "8092", 
      bandwidth: "10Mbps", 
      routing: "edge_2 edge_4 edge_3"},
    task_3: {
      task_id: "2 3", 
      source_node: "edge_3", 
      source_node_port: "8091", 
      previous_node: "edge_1 edge_2", 
      next_node: "", 
      next_node_ports: "", 
      bandwidth: "", 
      routing: ""},
  }
}
\end{lstlisting}
    \caption{Collaborative Task Scheduling Strategy}
    \label{f: scheduling_policy}
\end{figure}

The scheduler generates the collaborative task scheduling strategy, which decides where to allocate each task, how much the allocated bandwidth is, and the routing path together with the communication port for each data flow. As shown in Fig.~\ref{f: scheduling_policy}, the job shown in Fig.~\ref{f: application} is partitioned into $3$ tasks, where task$0$ and task$1$ are allocated to edge nodes $e1$ and $e2$, respectively. Task$2$ and task$3$ are both allocated to $e3$. The bandwidth of data flow $f_{01}$ and $f_{02}$ is restricted to $15$Mbps and $10$Mbps, respectively. The source node port and destination port of flow $f_{01}$ are set to be $8089$ and $8090$, respectively. The routing path of flow $f_{13}$ is determined as $\{e2, e3, e4\}$.   

Once the task scheduling strategy has been determined, they will be maintained by the compute controller and network controller, respectively, and sent to the edge nodes for execution. Specifically, the computation resource-related strategies, such as where to allocate the task and how many resources are assigned to the task, will be managed by Computer Controller, which interacts with the Kubelet on edge nodes to ensure the start, status monitoring, and stop of the containerized task. The networking resource-related strategies, such as port, bandwidth, and routing path of data flow, are managed by the network controller, which interacts with the network manager on edge nodes to ensure the communication and data transmission among edge nodes. The two controllers jointly manage the edge resources and ensure the correct execution of the collaborative task scheduling strategies with the coordination of the scheduler.

\subsection{Distributed Task Execution}
When the messenger receives the task execution policy, it will decompose the policies into computation-related and networking-related policies. The computation-related policies will be forwarded to and maintained by the Kubelet, while the networking-related ones will be forwarded to and maintained by the network manager. Kubeetl and network manager work together to ensure the proper execution of the assigned task.

\begin{figure}[t]
    \includegraphics[width=0.9\linewidth]{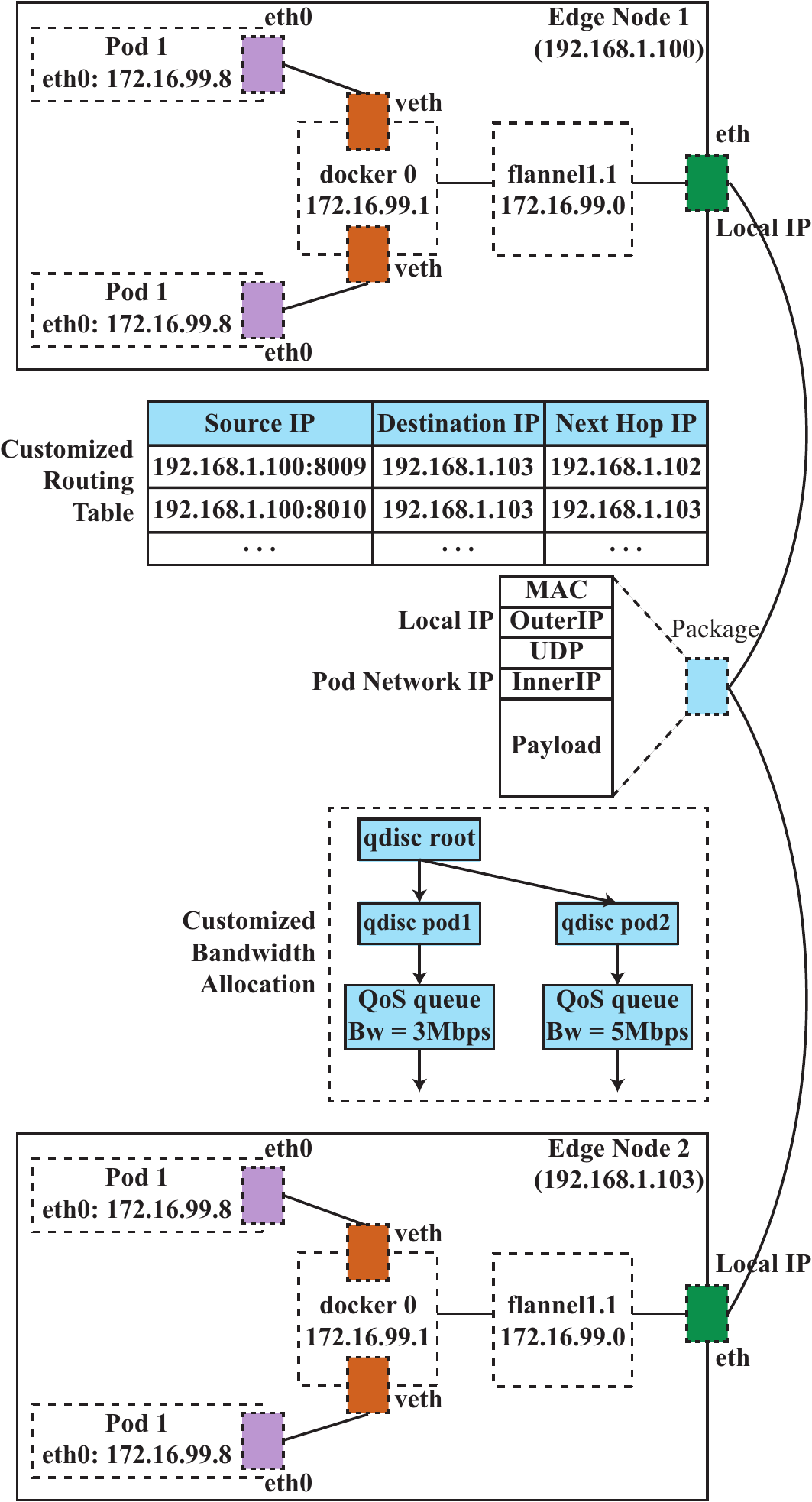} 
    \centering
    \caption{Bandwidth Allocation and Customized Routing of Network Manager}
    \label{network-manager}
\end{figure}

One important role of the network manager is to manage and orchestrate the networking resources. In this work, we are mainly concerned with the bandwidth allocation and customized routing of the cross-node data flows. For cross-node communication, Kubernetes usually adopts a flannel network \cite{dua2016learning}. As shown in Fig.~\ref{network-manager}, a data package from Pod1 to Pod3 will first be forward to docker0 and then to the flannel interface. The package will go through eth on edge node A and be sent to edge node B, where a reverse process will be performed to analyze the Internal IP of the package and route the package to the destination, i.e., Pod 3. To achieve the bandwidth allocation and customized routing of data flow, for each data flow in a scheduled job, the network manager will specify the \textit{\{source\_ip, source\_ip\_port, bandwidth\_limit, destination\_ip, destination\_ip\_port\}}, as shown in Fig.~\ref{f: scheduling_policy}. Through this information, the network manager leverages the Linux kernel functions, i.e., Traffic Control and Iproute \cite{hubert2002linux}, to shape the bandwidth between two edge nodes and customize routing for data packages. Traffic control creates Classful Queuing Disciplines (qdisc) to filter and redirect network packages to a particular quality-of-service queue before sending them out. The network manager also maintains the routing table of each assigned task. As shown in Fig.~\ref{network-manager}, the data package going through port $8009$ from edge node $1$ will be forwarded to another edge node rather than go directly to the destination, i.e., edge node $2$. Also, the bandwidth of data flow from Pod1 of edge node $1$ will be shaped to $3$Mbps.   

After the network configuration takes effect, the kublet will launch the pod according to the assigned computation-related policies, such as CPU and memory requests. The device monitor and task monitor will consistently and continuously monitor the status of the devices and the task.

\section{Collaborative Task Scheduling with Data Streaming Applications} \label{sec:Collaborative-Task-Scheduling}
In this section, we showcase the collaborative task scheduling of ENTS with representative data streaming applications, namely edge video analytics. We first introduce the system model. Then, we formulate a joint task allocation and flow scheduling problem for a single job scheduling and illustrate the proposed algorithms. On this basis, we further propose two online scheduling algorithms to schedule multiple continuous arriving jobs to maximize the average job throughput.

\subsection{System Model}
Edge video analytics \cite{zhang2019edge}\cite{yi2017lavea}\cite{zhang2022blockchain} is a killer application of edge computing. The network and application model used in formulating the problem is described as follows.

\subsubsection{Network Model}
The communication network is a mesh network of edge nodes connected using a multi-hop path. The network is modelled as an undirected graph $G = (V, E)$, where $V$ is the set of edge nodes, $V = \{j|1\leq j \leq M\}$, and $E$ is the set of links connecting different edge nodes, $E = \{l_{u,v}|u, v \in V\}$. Here, $M$ is the total number of edge nodes. The computing capacity, maximum resource and available resource of edge node $j$ is $PS_{j}$, $R_{max}^{j}$ and $R_{avail}^{j}$, respectively. The bandwidth of link $l$ is represented by $B_{l}$. The network can be heterogeneous in terms of the computation capacity of edge nodes and link bandwidth.

\subsubsection{Application Model} 
There will be multiple jobs submitted to the ENTS system by the edge nodes. Each job is modeled as a directed acyclic graph $J = (T, P)$, where $T$ is a set of dependent tasks and $P$ represents the set of dependencies between the tasks in the job. $Pd_{i}$ denotes the predecessor tasks of task $T_{i}$. The computation workload and resource demand of task $j$ is $C_{j}$ and $R_{req}^{j}$. The amount of dependent data between task $j$ and task $i$ is $D_{i,j}$. The input data source of job $J$ is assumed to be located at an edge node $s_{J}|s_{J} \in V $.

\subsection{Problem Formulation}
The objective of the single job scheduling is to maximize the throughput of the job by deciding where to allocate each task of the job, the routing path and bandwidth allocation of each data flow caused by the intermediate data transmission. If two dependent tasks are allocated to the same edge node, there will be no intermediate data transmission and thus no data flow. The joint task allocation and flow scheduling problem denoted as $P_{1}$ is formulated as follows.

\begin{equation}
\max \left\{TP = \frac{1}{t_{p}}\right\}
\end{equation}
\begin{equation} \label{eq:tp}
t_{p} = \max \left\{\max _{i \in T}(t_{comp}^{i}), \max _{i \in T, j \in Pd_{i}}(t_{comm}^{i,j})\right\}
\end{equation}
\begin{equation}\label{eq:comp}
t_{comp}^{i} = X_{i}^{u} \cdot \frac{C_{i}}{PS_{u}}
\end{equation}
\begin{equation}\label{eq:comm}
t_{comm}^{i,j} = X_{i}^{u} \cdot X_{j}^{v} \cdot \frac{D_{i,j}}{B_{l_{u,v}}}, j \in Pd_{i}
\end{equation}
\begin{equation}
    X_{i}^{u} \in\{0,1\}, \forall i, u
\end{equation}

Eq.~\ref{eq:comp} indicates the computation time of task $i$, where $X_{i}^{u}$ is a binary variable. $X_{i}^{u}$ equals to $1$ if task $i$ is allocated to edge node $u$, otherwise $X_{i}^{u}$ equals to $0$. Eq.~\ref{eq:comm} shows the transmission time of the intermediate data between dependent task $i$ and $j$. The throughput is $TP = \frac{1}{t_{p}}$, where $t_{p}$ is constraint by the maximum transmission and computation time as indicated by Eq.~\ref{eq:tp}. $P_{1}$ is a mixed Integrated Non-linear problem (MINLP), which is proven to be NP-hard in literature.

\subsection{Proposed Solution}
To solve the problem $P_{1}$, we decompose it into two sub-problems, i.e., allocate each task of the job $P_{2}$ and decide the routing path and bandwidth allocation of all the data flows $P_{3}$. To solve $P_{2}$, we use a greedy algorithm to allocate each task to the edge node, which can provide the least execution time, including the computation time and the dependent data transmission time. To solve $P_{3}$, we first relax it into a convex problem, which can be solved by convex optimizers, and then derive the solution for $P_{3}$.

\subsubsection{Solving Problem $P_{2}$} The algorithm to solve $P_{2}$ is shown in Algo.~\ref{algo:task-allocation}. For each task in the job, the algorithm traverses all the edge nodes with satisfied resource capacity and allocates the task to the edge node with the minimum execution time, including both computation time and intermediate data transmission time (Line 3-13). For calculating $t_{comm}^{i,j}$, we set the bandwidth between two edge nodes as the average bandwidth of all routing links. This is reasonable because the intermediate data flow can have multiple choices to avoid network congestion. Later, we will adjust the allocated bandwidth and the routing path of the data flows in a more fine-grained way in problem $P_{3}$.

\begin{algorithm}
\caption{Task Allocation}\label{algo:task-allocation}
\KwIn{network $G = (V, E)$, job $J = (T, P)$, }
\KwOut{the task allocation policy $T_{i,j}$, the data flows $FL$ }
Initialize $T_{i,j} \gets 0$ for all $i, j$\;
Query the available resource $R_{j}$ of all edge nodes\;
\For{task $T_{i}$ in job $J = (T, P)$}
{
    \For{edge node $j$ in network $G = (V, E)$}
    {
        \If{$R_{avail}^{j} > R_{req}^{i}$}
        {
            Calculate the computation time $t_{comp}^{i} = C_{i} \div PS_{j}$\;
            Calculate the intermediate data transmission time $t_{comm}^{i} = \max t_{comm}^{i,j}$ using Eq. (4)\;
            Calculate the execution time $t_{exec}^{j} = t_{comp}^{j} + t_{comm}^{j}$\;
        }
    }
    Allocate task $T_{i}$ to node $j^{*} = min_{J} \{t_{exec}^{j}\}$\;
    $T_{i,j^{*}} \gets 1$\;
    Update $R_{j*}$ for node $j^{*}$\;
}
Calculate data flow $f_{i} = <source, destination, datasize>$ with $T_{i,j}$\;
Add $f_{i}$ to data flows $FL$\;
\Return $T_{i,j}$, $FL$
\end{algorithm}

\subsubsection{Solving Problem $P_{3}$} After solving $P_{2}$, we get the data flows $FL$, where we can know the number of data flows $Nf$, the source, destination, and data volume of each data flow $f_{i}$. We then solve $P_{3}$ to decide the routing path and bandwidth allocation of each data flow. The $P_{3}$ is formulated as follows.

\begin{equation}
\min \max _{i=1, \ldots, Nf}\left\{\frac{V_{i}}{b_{i}}\right\}
\end{equation}

\begin{equation}\label{eq:bandwidth-constraint}
    \sum_{i} \sum_{k: l \in P_{i}^{k}} b_{i} y_{i}^{k} \leq B_{l}, \forall l
\end{equation}
\begin{equation}\label{eq:sum_y}
    \sum_{k} y_{i}^{k}=1, \quad \forall i
\end{equation}
\begin{equation}\label{eq:binary_y}
    y_{i}^{k} \in\{0,1\}, \forall i, k
\end{equation}
where $V_{i}$ is the size of flow $f_{i}$ and $b_{i}$ is the bandwidth allocated to flow $f_{i}$. $P_{i}^{k}$ is the collection of all the possible routing paths of flow $f_{i}$. $y_{i}^{k}$ is a binary variable. $y_{i}^{k}$ equals to $1$ if flow $f_{i}$ chooses the $k^{th}$ routing path of $P_{i}^{k}$. Note that Eq.~\ref{eq:bandwidth-constraint} indicates that the sum of allocated bandwidth of all data flows going through link $l$ cannot exceed its capacity. Eq.~\ref{eq:sum_y} and Eq.~\ref{eq:binary_y} ensure that a data flow can only choose one routing path.

The problem $P_{3}$ is still a MINLP problem. Therefore, we resort to relaxing the integer variable $y_{i}^{k}$ to a real variable $y_{i}^{k} \geq 0$. We name the relaxed problem $P_{3}-\textsc{Relax}$. Due to the existence of term $b_{i} \cdot y_{i}^{k}$, the $P_{3}-\textsc{Relax}$ problem is still a non-linear programming problem which is hard to solve. In the following, we transform the $P_{3}-\textsc{Relax}$ problem into an equivalent convex optimization problem. 

\subsubsection{An Equivalent Convex Problem} First, we introduce an variable $TH$ such that $TH = \max \limits_{i=1, \ldots, Nf}\left\{\frac{V_{i}}{b_{i}}\right\}$. Furthermore, we introduce another variable $q_{i}$ such that $q_{i} = TH \cdot b_{i}$, and variable $m_{i}^{k} = q_{i} \cdot y_{i}^{k}$. Then, the equivalent problem $P_{3}-\textsc{Relax-Cvx}$ is formulated below.
\begin{equation}
    \min TH
\end{equation}

\begin{equation}
    \sum_{i} \sum_{k: l \in P_{i}^{k}} m_{i}^{k} \leq B_{l} \cdot TH, \forall l 
\end{equation}
\begin{equation}
    \sum_{k} m_{i}^{k}=q_{i}, \quad \forall i 
\end{equation}
\begin{equation}
    m_{i}^{k} \geq 0, \forall i, k
\end{equation}
\begin{equation}
    q_{i} \geq V_{i}, \forall i
\end{equation}
All constraint in the $P_{3}-\textsc{Relax-Cvx}$ is affine, and the objective function is convex. Therefore, the $P_{3}-\textsc{Relax-Cvx}$ problem is a convex optimization problem which can be solved using convex optimizers \cite{boyd2004convex}. 

However, since we relax the binary integer constraint, the solution may be that some $y_{i}^{k}$ are decimal factions. To solve the problem, we route the $i^{th}$ data flow to a path $k^{*}$ such that $m_{i}^{k^{*}} = \max_{k} m_{i}^{k}$. When the routing path is determined, the optimal bandwidth allocation policies is given by 
\begin{equation}\label{eq:b*}
    b_{i}^{*} = \min \left\{\frac{V_{i}}{\sum_{i} \sum_{k: l \in P_{i}^{k^{*}}} V_{i} y_{i}^{k^{*}}}\right\}, l \in P_{i}^{k^{*}}
\end{equation}

The algorithm to solve $P_{3}$ is shown in Algo.~\ref{algo:JoRBA}.

\begin{algorithm}[t]
\caption{Joint Routing and Bandwidth Allocation (JRBA)}\label{algo:JoRBA}
\KwIn{network $G = (V, E)$, data flows $FL$, }
\KwOut{the routing policy $y_{i}^{k}$, the bandwidth allocation policy $b_{i}$, and job throughput $JTH$}
Solve $P_{3}-\textsc{Relax-Cvx}$ and get $\{T^{*}, q_{i}^{*}, m^{k^{*}}_{i}\}$\;
\For{flow $f_{i}$ in $FL$}
{
    Initialize $y_{i}^{k} \gets 0$ for all $k$\;
    $k^{*} \gets arg_{k}\max m_{i}^{k}$\;
    $y_{i}^{k^{*}} \gets 1$\;
}
Calculate $b_{i}^{*}$ using Eq.~\ref{eq:b*}\;
Update $B_{l}$ according to $y_{i}^{k^{*}}$, $b_{i}^{*}$\;
$JTH \gets  \max _{i=1, \ldots, N}\left\{\frac{V_{i}}{b_{i}}\right\}$\;
\Return $y_{i}^{k}$, $b_{i}$, $JTH$
\end{algorithm}

\subsection{Online Scheduling}
Algo.~\ref{algo:task-allocation} and Algo.~\ref{algo:JoRBA} study the task scheduling for one job. However, in a practical ENTS system, jobs constantly arrive and share the resource in the network. Our goal is to maximize the average job throughput. Motivated by this, we propose two online scheduling algorithms, which run in the ENTS online scheduler and periodically schedule all arrived jobs.

The online scheduler maintains two job queues: 1) a queue of jobs that are running, denoted by $Q_{run}$, and 2) a queue of jobs that are waiting to be scheduled, denoted by $Q_{wait}$. The two online scheduling algorithms are: 1) schedule the job in $Q_{wait}$ one by one, and 2) schedule the job in $Q_{wait}$ one by one but readjust the routing and bandwidth sharing strategy by considering all the existing and coming data flows in the edge network. 

The first algorithm (OTFS) is shown in Algo.~\ref{algo:online}. For each job in the queue $Q_{wait}$, the algorithm first sorts the job in descending order of waiting time and schedules the jobs in sequence (Line 6-9). During scheduling, the algorithm calls the procedure Task Allocation (Algo.~\ref{algo:task-allocation}) and JRBA (Algo.~\ref{algo:JoRBA}) in turn (Line 9-13). 

The second algorithm (OTFA) is shown in Algo.~\ref{algo:OSJFA}. Different from OTFS, which makes task scheduling decisions based on the current status of the computation and networking resource in the edge network, OTFA jointly manages the existing data flows and the coming data flows. It first allocates the computation resources for arriving jobs and then readjusts the networking resources for all data flows (Line 10-15). 

\begin{algorithm}[t]
\caption{OTFS: Online Task Allocation and Flow Scheduling}\label{algo:online}
\KwIn{current time $curT$, network $G = (V, E)$, $Q_{wait}$}
$J_{finish} \gets$ all jobs finishing at $curT$\;
\If{$J_{finish} \neq \emptyset$}
{
    Release all computing resource and bandwidth allocated to $J_{finish}$\;
    Update $R_{j}$ and $B_{l}$ for network\;
}
\If{there are jobs arriving at $curT$}
{
    Add jobs arriving at $curT$ to $Q_{wait}$\;
}
Sort $Q_{wait}$ in descending order of waiting time\;
\For{job $J_{i}$ in $Q_{wait}$}
{
    Call the Task Allocation procedure to get $\{T_{i,j}, FL\}$\;
    Call the JRBA procedure\;
}
\end{algorithm}

\begin{algorithm}[t]
\caption{OTFA: Online Scheduling Task Allocation Joint Flow Adjustment}\label{algo:OSJFA}
\KwIn{current time $curT$, network $G = (V, E)$, $Q_{wait}$, $Q_{run}$}
$J_{finish} \gets$ all jobs finishing at $curT$\;
\If{$J_{finish} \neq \emptyset$}
{
    Release all computing resource and bandwidth allocated to $J_{finish}$\;
    Update $R_{j}$ and $B_{l}$ for network\;
}
\If{there are jobs arriving at $curT$}
{
    Add jobs arriving at $curT$ to $Q_{wait}$\;
}
Sort $Q_{wait}$ in descending order of waiting time\;
\For{job $J_{i}$ in $Q_{wait}$}
{
    Call the Task Allocation procedure to get $\{T_{i,j}, FL\}$\;
}
Release all bandwidth allocated to data flows $FL_{run}$ in $Q_{run}$\;
Add $FL$ to $FL_{run}$\;
Call the procedure JRBA with $FL_{run}$\;
\end{algorithm}

\section{Experimental Results}\label{sec:exp-results}

\begin{figure}[t]
    \centering
    \includegraphics[width=0.77\linewidth]{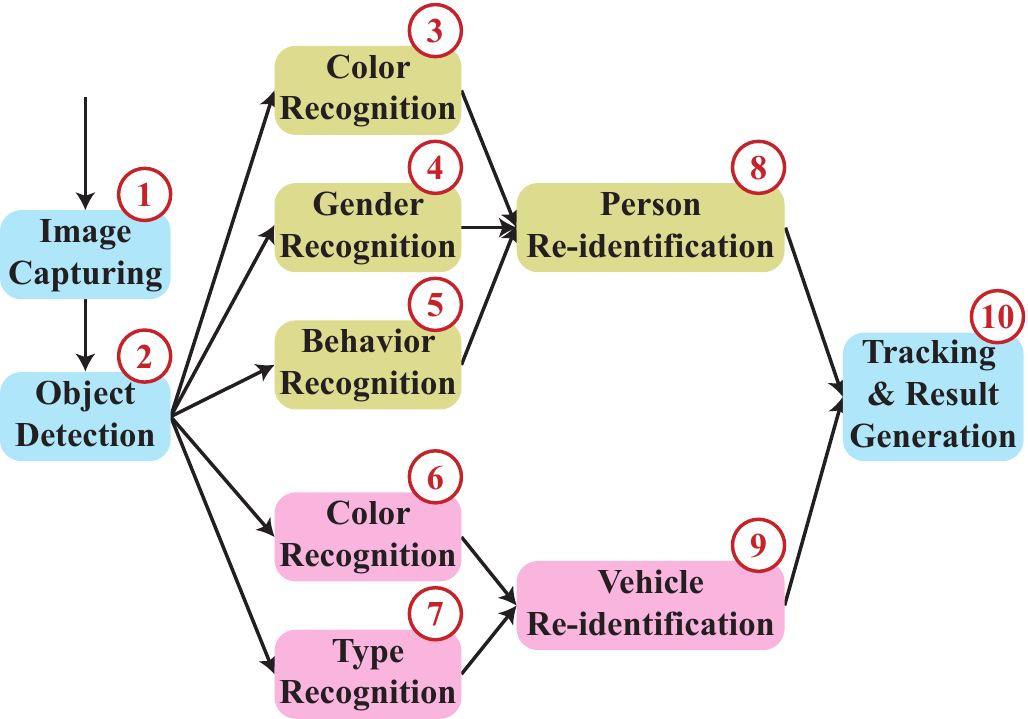}
    \caption{Application Graph of Object Attributes Recognition}
    \label{f:application}
\end{figure}

\subsection{Experimental Setup}
\subsubsection{Benchmarks} To evaluate the ENTS system, we use a real-world live video analytics application, i.e., object attribute recognition \cite{zeng2020distream}, which is extensively used in surveillance of public safety. The application graph is shown in Fig.~\ref{f:application}, where we have $10$ functional modules. For modules $2$ to $9$, each of them is implemented with a computing-extensive and resource-greedy DNN model \cite{jacob2018quantization}\cite{yang2021tree}. The application takes the surveillance video as input and recognizes the attributes of pedestrians and vehicles in the video, such as the color of cloth, gender of pedestrians, and type of vehicles. Specifically, we use MobileNet-V2 \cite{sandler2018mobilenetv2} as the backbone network for object detection in module $2$. For attribute recognition and object re-identification, i.e., module $3-9$, we use Resnet-50 \cite{hermans2017defense} as the backbone network. We use the Kalman filter to track the objects in module $10$. The resolution of the video is 1920x1080 with $30$fps and the size of each video frame is about $6$MB. The application is implemented with Python. 

\subsubsection{Baselines} We compared the proposed method with three state-of-the-art baselines as follows.
\begin{itemize}
    \item \textit{LeastRequestPriority (LR).} It schedules the whole job to the edge node with the least resource consumption. The LR policy is frequently used in Kubernetes.
    \item \textit{BalancedResourceAllocation (BR).} It schedules the whole job to the edge node, which can balance the resource consumption among the edge nodes. BR is used in Kubernetes to achieve workload balancing. 
    \item \textit{Task Partition (TP).} It partitions the job and schedules each task to the edge nodes with the least execution time, including the transmission time and the computation time. We adopt the default shortest path to transfer the intermediate data. When multiple data flows go through the same link, all flows equally share the link bandwidth.
\end{itemize}

\subsubsection{Metrics} We employ two metrics as follows.
\begin{itemize}
    \item \textit{Average Job Throughput.} It is the average throughput of all submitted jobs. It is an important metric to measure the performance of the scheduling algorithms.
    \item \textit{Average Waiting Time.} It is the average waiting time of all submitted jobs, i.e., the time from the job submitted to the job scheduled. It is a metric reflecting the effectiveness of the scheduler and system overhead.
\end{itemize}

\begin{figure}[t]
    \centering
    \includegraphics[width=.8\linewidth]{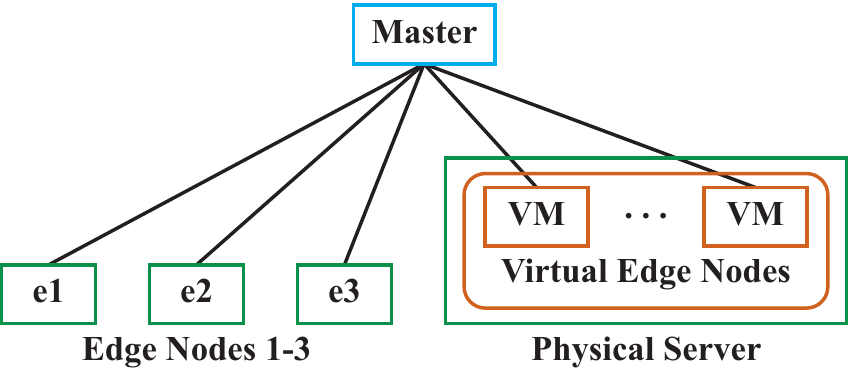}
    \caption{Test Environment of ENTS}
    \label{f:testbed}
\end{figure}

\subsubsection{Testbed Implementation}
To test the system on a large scale geo-distributed edge environment, we developed a hybrid testbed with both physical and virtual edge nodes, as shown in Fig.~\ref{f:testbed}. We use virtual machines to emulate virtual edge nodes. While numerous virtual edge nodes enable us to test in a large-scale and network-flexible testing environment, the incorporation of physical nodes guarantees the fidelity of the testbed. We leverage Linux Traffic Control to configure the network topology and bandwidth among the edge nodes. We vary the network link bandwidth, e.g., from $1Mbps$ to $10Mbps$, to emulate the physical distance among edge nodes. The intuition is that the bandwidth should be low if two nodes are far away. Similar idea is also adopted in \cite{krajsa2011rtt}.

Specifically, we randomly generate the network connection among edge nodes with the average node degree as $3$. We also enable routing and forwarding on each node so that each node is both a compute node and a router. We use $4$ raspberry pi, $2$ Nvidia Jetson Nano, and $2$ Nvidia Jetson Xavier NX to represent physical edge nodes. A PC equipped with four Intel Cores i9-7100U with $20$GB RAM to act as the master node to manage the edge nodes. Two servers are leveraged to host virtual machines acting as virtual edge nodes. One is equipped with Intel(R) Xeon(R) Gold 6128 CPU with $192$GB Memory, another is Intel(R) Core(TM) i9-10900F CPU with $64$GB memory. The specifications of the physical devices are shown in TAB.~\ref{t:physical-devices}.

\subsection{Results and Analysis}
We test the performance of the ENTS system and the proposed online scheduling algorithms under various situations.

\subsubsection{Effects of Number of Edge Nodes}
We evaluate the influence of the number of edge nodes on the average job throughput and average waiting time to test the scalability of ENTS. In this experiment, a total of 50 jobs are submitted by the edge nodes to the master with the arriving rate following a Poisson distribution with $\lambda = 0.5/second$. 

\begin{figure*}[t]
    \centering
    \includegraphics[width=.84\linewidth]{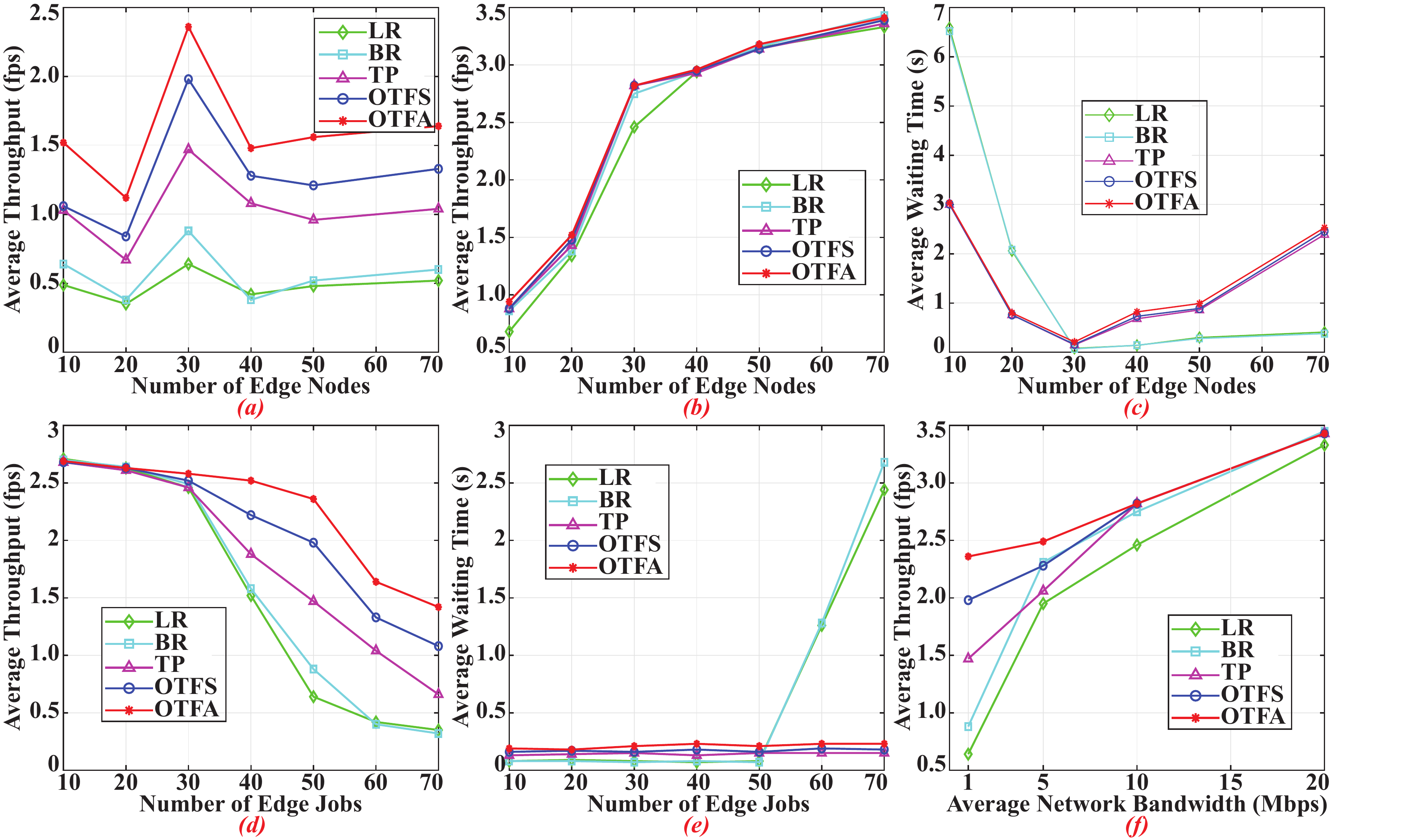}
    \caption{a) Impact of the number of edge nodes on average throughput with average bandwidth $1$Mbps. b) Impact of the number of edge nodes on average throughput with average bandwidth $10$Mbps. c) Impact of the number of edge nodes on average waiting time. d) Impact of the number of submitted jobs on average throughput. e) Impact of the number of submitted jobs on average waiting time. f) Impact of average bandwidth on average throughput.}
    \label{fig:exp-results}
\end{figure*}

As shown in Fig.~\ref{fig:exp-results}(a), TR, OTFS, and OTFA perform much better than LR and BR, with higher average throughput. The average throughput of LR and BR does not exceed $1$. It is because LR and BR do not partition the job, which leads to the transmission of source video data over a low-bandwidth edge network. It becomes the bottleneck of the job throughput. Unlike LR and BR, the other three methods, i.e., TP, OTFS, and OTFA, partition the job and enable distributed job execution, avoiding raw data transmission. OTFA performs best with the highest throughput among TP, OTFS, and OTFA. TP shares the bandwidth equally and assigns the shortest routing path for network flows, which usually leads to traffic congestion when multiple data flows pass through the same network link. Instead, OTFS and OTFA optimize the networking resources by enabling optimal bandwidth sharing and routing path selection concerning the end-to-end job throughput. OTFA goes further. It considers all the available data flows in the network, which can improve the average job throughput compared to OTFS.

\begin{table}[t]
    \centering
    \caption{Specifications of Physical Devices}
    \label{t:physical-devices} 
     \begin{tabular}{cccc}
        \toprule
        Name & CPU & Memory & Performance \\
        \toprule
        Raspberry Pi & $1$ core & $1$GB & Low  \\
        Jetson Nano & $6$ cores & $4$GB & Low  \\ 
        Jetson Xavier NX & $6$ cores & $8$GB & Medium  \\
        Edge Server-1 & $64$ cores & $64$GB & High \\
        Edge Server-2 & $128$ cores & $192$GB & High  \\
        \bottomrule
     \end{tabular}
\end{table}

We also observe that the average throughput does not show a linear growth with an increasing number of edge nodes. Generally, when the number of edge nodes increases, the network will have more resources and higher job throughput. However, the throughput decreases slightly when the number of edge nodes increases from $10$ to $20$ and $30$ to $40$. It is because of the limited network bandwidth, i.e., $1$Mbps with a variance of $0.3$ in our experiment. When the number of edge nodes increases, the number of hops and network links between two edge nodes also increases, resulting in more bottleneck communication paths. As shown in Fig.~\ref{fig:exp-results}(b), when the average bandwidth of the edge network becomes $10$Mbps, such fluctuation of the average throughput will no longer exist. More specifically, it shows a linear growth as expected. It is because the network bandwidth is not the bottleneck anymore, and there are fewer bottleneck communication paths.

Fig.~\ref{fig:exp-results}(c) depicts the influence of the number of edge nodes on the waiting time. When the number of the edge nodes is below $30$, the average waiting time of TP, OTFS, and OTFA is much smaller than that of LR and BR. The reason is that the former scheduling policies partition the job and allocate the task into edge nodes with less abundant resources, improving resource utilization and the number of jobs executable among the geo-distributed edge nodes. When the number of edge nodes is above $30$, the total resource is sufficient, where the average waiting time is dominated by the running efficiency of the scheduling algorithms. Compared with the LR and BR algorithms, TP, OTFS, and OTFA are required to traverse all the edge nodes for each task and solve the formulated optimization problem, which increases the average waiting time. However, we observe that when the number of edge nodes is below $50$, the average waiting time is no more than $1$ second, and about $2.5$ second when the number of edge nodes is $70$, which is still at a low level.

\subsubsection{Effects of Number of Submitted Jobs}
We evaluate the performance of the average job throughput and wait time with a changing number of submitted jobs. We set the average bandwidth as $1$Mbps with a variance of $0.3$. The number of edge nodes is $30$. The arriving rate of the submitted jobs follows a Poisson distributed with $\lambda = 0.5/second$. 

As shown in Fig.~\ref{fig:exp-results}(d), when the number of submitted jobs is no more than $30$, our method performs similarly to the baseline. In such cases, the edge resources are relatively abundant, and the proposed methods, i.e., OTFS and OTFA, tend to yield similar decisions compared with the baseline methods. However, when there are more jobs, the average throughput of LR and BR declines dramatically. It is because multiple jobs compete for limited networking and computation resources. Without partitioning the submitted jobs and optimizing the bandwidth allocation and routing path of flows, LR and BR easily suffer from network congestion and fragmented computation resource usage, degrading the average job throughput significantly. OTFA performs the best. Compared to TR and OTFS, OTFA considers optimal bandwidth sharing and routing path for incoming in addition to existing data flows, which can further improve the averaging job throughput with better resource utilization when there are more jobs.

Fig.~\ref{fig:exp-results}(e) depicts similar trends concerning the performance in average waiting time. When the number of submitted jobs is below $50$, the average waiting time for all the mentioned methods is low, i.e., no larger than $0.5$ without apparent fluctuation. We can also see that the waiting time of LR and BR is shorter than that of TP, OTFS, and OTFA. It is because the latter three approaches have to traverse all the edge nodes for each task, which leads to more waiting time for scheduling jobs. When the number of submitted jobs exceeds $50$, TP, OTFS, and OTFA show consistent average waiting times while the performance of LR and BR increases significantly. The reason is that there are no available resources to schedule the new-coming jobs. The rest of the jobs are required to wait in the job queue, which results in an increased average waiting time. Compared to TR, OTFS, and OTFA, the other two methods, i.e., BR and LR, do not partition the submitted job, which may easily lead to fragmented resource consumption and thus serve fewer jobs.
 
\subsubsection{Effects of Average Bandwidth}
We also evaluate the performance of the average job throughput with the variance of the average bandwidth of the edge network. We set the number of edge nodes as $30$ in this experiment and the number of submitted jobs as $50$ with the arriving rate following a Poisson distributed with $\lambda = 0.5/second$.

As shown in Fig.~\ref{fig:exp-results}(f), the average throughput of all the methods increases with the average bandwidth. More specifically, when the average bandwidth of the edge network is no more than $5$Mbps, OTFA outperforms other methods significantly because it jointly considers and optimizes the data locality, the networking, and computing resources of edge nodes. However, when the average bandwidth is above $10$Mbps, baselines and proposed methods tend to have similar performance. It is because the bandwidth is relatively abundant now. However, OTFS and OTFA are slightly better than LR and BR, as they optimize the bandwidth allocation and routing selection for data flows in the edge network. BR outperforms LR as it aims to achieve balanced resource consumption, enabling the powerful edge nodes to service more jobs.

In a nutshell, we evaluated and compared the performance of ENTS with the state-of-the-art and proposed online algorithms for scheduling streaming jobs. Benefiting from the ability to consider task dependencies and jointly optimize the limited coupled computation and networking resources, ENTS achieves a $43\%-220\%$ improvement in average throughput. Although the proposed solutions introduce additional overhead in making the scheduling strategies, they can serve more jobs when resources of the edge network are limited, which leads to less averaging waiting time.

\section{Related Work}\label{sec:related-work}

\textit{Container scheduler.} The default scheduler of Kubernetes (K8S) \cite{luksa2017kubernetes}  is an online scheduler that implements a greedy multi-criteria decision-making (MCDM) algorithm. MCDM scores the available nodes with pre-defined rules and selects the highest scoring node for scheduling. This scheduling algorithm performs well in the cloud environment. However, it lacks features for container scheduling in the edge environment, such as limited network connections and geo-distributed and resource-constraint edge nodes. Furthermore, it is not performance-aware. There are several attempts to tailor the Kubernetes for the edge.
Regarding the resource-constraint edge environment, MicroK8s and K3s \cite{kjorveziroski2022kubernetes} aims to simplify K8S and provide lightweight K8S distribution. KubeEdge \cite{xiong2018extend} and OpenYurt extend the K8S capability to the edge by enabling the virtual network connection between edge servers and VMs in the cloud. However, those solutions do not change the core idea of task scheduling of Kubernetes. They are not application performance sensitive.

Some work tries to improve the scheduling policies for performance-sensitive edge applications. Santos et al. \cite{santos2019towards} tried to extend the default task scheduling strategies in Kubernetes with the ability to sense the network status. They consider the round trip time information of candidate nodes to minimize the overall response time of an application to be deployed. Rossi et al. \cite{rossi2020geo} designed a customized scheduler leveraging the Monitor, Analyze, Planning, Execute (MAPE) pattern to deploy applications in a geo-distributed environment. Wojciechowski et al. \cite{wojciechowski2021netmarks} proposed NetMARKS, fulfilling the Kubernetes scheduler with the network-aware feature. It uses Istio service mesh to collect network metrics, facilitating scheduling pods on a server and its neighbors and encouraging co-locating pods \cite{larsson2020impact}. Though these works consider the network latency between edge nodes, it neglects the heterogeneous computing capability of edge nodes and the locality of data sources. Besides, they do not orchestrate the networking resource, such as bandwidth allocation and customized routing of data flows.

\textit{Task scheduling in cloud-edge infrastructure.} Many works consider dispatching streaming tasks among heterogeneous edge servers and cloud to minimize the average task completion time \cite{meng2019dedas,han2019ondisc}. However, they only consider the independent tasks while neglecting the dependency among tasks. Concerning dependent tasks, Sundar et al. \cite{sundar2018offloading} proposed a heuristic algorithm for scheduling dependent tasks in a generic cloud computing system by greedily optimizing the scheduling of each task subject to its time constraint. Wang et al. \cite{wang2021dependent} developed a deep reinforcement learning-based task offloading scheme, which leverages the off-policy reinforcement learning algorithm with a sequence-to-sequence neural network to capture the task dependency of applications. Nevertheless, they fail to consider the orchestration of the network flows \cite{liu2020dependency}, which necessarily results in network congestion and prolonged task completion time. Although there are some works \cite{sahni2018data,sahni2020multi} optimizing the average task completion time and jointly considering the task allocation and flow scheduling, they do not optimize the application throughput and lack real-world system implementation. 

To summarize, different from existing works, we jointly consider the data, computing, and networking resource to maximize the throughput of stream applications and proposed and developed a holistic system to enable application development, online scheduling, and distributed task execution.

\section{Conclusion and Future Work}\label{sec:conclusion}
In this work, we designed and developed ENTS, the first edge-native task scheduling system, to manage geo-distributed and heterogeneous edge resources in collaborative edge computing. ENTS extends Kubernetes with the ability to jointly orchestrate computation and networking resources to optimize the application performance. ENTS comprehensively considers both the application characteristics and edge resource status. We show the superiority of ENTS with a case study on data streaming applications, in which we formulate a joint task allocation and flow scheduling problem and propose two online scheduling algorithms. Experiments on an object attribute recognition application on a large number of edge nodes show ENTS achieves improved performance.  

In the future, we will improve the work from two aspects as follows. On the one hand, we will develop more advanced algorithms for collaborative task scheduling. Current algorithms do not allocate resources for tasks, such as how much memory and CPU periods should be allocated to the containerized tasks. However, regarding optimization of the overall resource usage, we have to jointly consider the task partition and allocation, computing resource allocation, and networking resource allocation. On the other hand, we will integrate software-defined networking (SDN) into the network controller. We use the Linux kernel functions, i.e., Iproute and Traffic control, to achieve networking resource management for the network manager. The objective is consistent with SDN, which provides programming interfaces for conveniently orchestrating networking resources. Many works \cite{baktir2017can,li2018adaptive,wang2019software} are exploring integrating SDN with edge computing to facilitate the management of various edge nodes.

\section{Acknowledgement}
This work was supported by the Research Institute for Artificial Intelligence of Things, The Hong Kong Polytechnic University, HK RGC Research Impact Fund No. R5060-19, and General Research Fund No. PolyU 15220020.


\bibliographystyle{IEEEtran}
\bibliography{main.bib}

\begin{thebibliography}{10}
\providecommand{\url}[1]{#1}
\csname url@samestyle\endcsname
\providecommand{\newblock}{\relax}
\providecommand{\bibinfo}[2]{#2}
\providecommand{\BIBentrySTDinterwordspacing}{\spaceskip=0pt\relax}
\providecommand{\BIBentryALTinterwordstretchfactor}{4}
\providecommand{\BIBentryALTinterwordspacing}{\spaceskip=\fontdimen2\font plus
\BIBentryALTinterwordstretchfactor\fontdimen3\font minus
  \fontdimen4\font\relax}
\providecommand{\BIBforeignlanguage}[2]{{%
\expandafter\ifx\csname l@#1\endcsname\relax
\typeout{** WARNING: IEEEtran.bst: No hyphenation pattern has been}%
\typeout{** loaded for the language `#1'. Using the pattern for}%
\typeout{** the default language instead.}%
\else
\language=\csname l@#1\endcsname
\fi
#2}}
\providecommand{\BIBdecl}{\relax}
\BIBdecl

\bibitem{shi2016edge}
W.~Shi, J.~Cao, Q.~Zhang, Y.~Li, and L.~Xu, ``Edge computing: Vision and
  challenges,'' \emph{IEEE Internet of Things Journal}, vol.~3, no.~5, pp.
  637--646, 2016.

\bibitem{mao2017survey}
Y.~Mao, C.~You, J.~Zhang, K.~Huang, and K.~B. Letaief, ``A survey on mobile
  edge computing: The communication perspective,'' \emph{IEEE Communications
  Surveys \& Tutorials}, vol.~19, no.~4, pp. 2322--2358, 2017.

\bibitem{chen2019deep}
J.~Chen and X.~Ran, ``Deep learning with edge computing: A review,''
  \emph{Proceedings of the IEEE}, vol. 107, no.~8, pp. 1655--1674, 2019.

\bibitem{zhang2017towards}
W.~Zhang, J.~Chen, Y.~Zhang, and D.~Raychaudhuri, ``Towards efficient edge
  cloud augmentation for virtual reality mmogs,'' in \emph{ACM/IEEE Symposium
  on Edge Computing}, 2017, pp. 1--14.

\bibitem{liu2019edge}
S.~Liu, L.~Liu, J.~Tang, B.~Yu, Y.~Wang, and W.~Shi, ``Edge computing for
  autonomous driving: Opportunities and challenges,'' \emph{Proceedings of the
  IEEE}, vol. 107, no.~8, pp. 1697--1716, 2019.

\bibitem{sacco2020edge}
A.~Sacco, F.~Esposito, G.~Marchetto, G.~Kolar, and K.~Schwetye, ``On edge
  computing for remote pathology consultations and computations,'' \emph{IEEE
  Journal of Biomedical and Health Informatics}, vol.~24, no.~9, pp.
  2523--2534, 2020.

\bibitem{zhang2022eaas}
M.~Zhang, J.~Cao, Y.~Sahni, Q.~Chen, S.~Jiang, and T.~Wu, ``Eaas: A
  service-oriented edge computing framework towards distributed intelligence,''
  \emph{arXiv preprint arXiv:2209.06613}, 2022.

\bibitem{ning2018green}
Z.~Ning, X.~Kong, F.~Xia, W.~Hou, and X.~Wang, ``Green and sustainable cloud of
  things: Enabling collaborative edge computing,'' \emph{IEEE Communications
  Magazine}, vol.~57, no.~1, pp. 72--78, 2018.

\bibitem{meng2019dedas}
J.~Meng, H.~Tan, C.~Xu, W.~Cao, L.~Liu, and B.~Li, ``Dedas: Online task
  dispatching and scheduling with bandwidth constraint in edge computing,'' in
  \emph{IEEE Conference on Computer Communications}, 2019, pp. 2287--2295.

\bibitem{zhang2021joint}
J.~Zhang, X.~Zhou, T.~Ge, X.~Wang, and T.~Hwang, ``Joint task scheduling and
  containerizing for efficient edge computing,'' \emph{IEEE Transactions on
  Parallel and Distributed Systems}, vol.~32, no.~8, pp. 2086--2100, 2021.

\bibitem{soppelsa2016native}
F.~Soppelsa and C.~Kaewkasi, \emph{Native docker clustering with swarm}.\hskip
  1em plus 0.5em minus 0.4em\relax Packt Publishing, 2016.

\bibitem{luksa2017kubernetes}
M.~Luksa, \emph{Kubernetes in action}.\hskip 1em plus 0.5em minus 0.4em\relax
  Simon and Schuster, 2017.

\bibitem{hindman2011mesos}
B.~Hindman, A.~Konwinski, M.~Zaharia, A.~Ghodsi, A.~D. Joseph, R.~Katz,
  S.~Shenker, and I.~Stoica, ``Mesos: A platform for $\{$Fine-Grained$\}$
  resource sharing in the data center,'' in \emph{USENIX Symposium on Networked
  Systems Design and Implementation}, 2011.

\bibitem{burns2016borg}
B.~Burns, B.~Grant, D.~Oppenheimer, E.~Brewer, and J.~Wilkes, ``Borg, omega,
  and kubernetes,'' \emph{Communications of the ACM}, vol.~59, no.~5, pp.
  50--57, 2016.

\bibitem{brewer2015kubernetes}
E.~A. Brewer, ``Kubernetes and the path to cloud native,'' in \emph{ACM
  Symposium on Cloud Computing}, 2015, pp. 167--167.

\bibitem{han2021tailored}
Y.~Han, S.~Shen, X.~Wang, S.~Wang, and V.~C. Leung, ``Tailored learning-based
  scheduling for kubernetes-oriented edge-cloud system,'' in \emph{IEEE
  Conference on Computer Communications}, 2021, pp. 1--10.

\bibitem{rossi2020geo}
F.~Rossi, V.~Cardellini, F.~L. Presti, and M.~Nardelli, ``Geo-distributed
  efficient deployment of containers with kubernetes,'' \emph{Computer
  Communications}, vol. 159, pp. 161--174, 2020.

\bibitem{wojciechowski2021netmarks}
{\L}.~Wojciechowski, K.~Opasiak, J.~Latusek, M.~Wereski, V.~Morales, T.~Kim,
  and M.~Hong, ``Netmarks: Network metrics-aware kubernetes scheduler powered
  by service mesh,'' in \emph{IEEE Conference on Computer Communications},
  2021, pp. 1--9.

\bibitem{liu2020dependency}
Y.~Liu, S.~Wang, Q.~Zhao, S.~Du, A.~Zhou, X.~Ma, and F.~Yang,
  ``Dependency-aware task scheduling in vehicular edge computing,'' \emph{IEEE
  Internet of Things Journal}, vol.~7, no.~6, pp. 4961--4971, 2020.

\bibitem{johnston2004advances}
W.~M. Johnston, J.~P. Hanna, and R.~J. Millar, ``Advances in dataflow
  programming languages,'' \emph{ACM Computing Surveys}, vol.~36, no.~1, pp.
  1--34, 2004.

\bibitem{zhang2019edge}
Q.~Zhang, H.~Sun, X.~Wu, and H.~Zhong, ``Edge video analytics for public
  safety: A review,'' \emph{Proceedings of the IEEE}, vol. 107, no.~8, pp.
  1675--1696, 2019.

\bibitem{kwon2013mantis}
Y.~Kwon, S.~Lee, H.~Yi, D.~Kwon, S.~Yang, B.-G. Chun, L.~Huang, P.~Maniatis,
  M.~Naik, and Y.~Paek, ``Mantis: Automatic performance prediction for
  smartphone applications,'' in \emph{USENIX Annual Technical Conference},
  2013, pp. 297--308.

\bibitem{pham2017predicting}
T.-P. Pham, J.~J. Durillo, and T.~Fahringer, ``Predicting workflow task
  execution time in the cloud using a two-stage machine learning approach,''
  \emph{IEEE Transactions on Cloud Computing}, vol.~8, no.~1, pp. 256--268,
  2017.

\bibitem{dua2016learning}
R.~Dua, V.~Kohli, and S.~K. Konduri, \emph{Learning Docker Networking}.\hskip
  1em plus 0.5em minus 0.4em\relax Packt Publishing, 2016.

\bibitem{hubert2002linux}
B.~Hubert \emph{et~al.}, ``Linux advanced routing \& traffic control howto,''
  \emph{Netherlabs BV}, vol.~1, pp. 99--107, 2002.

\bibitem{yi2017lavea}
S.~Yi, Z.~Hao, Q.~Zhang, Q.~Zhang, W.~Shi, and Q.~Li, ``Lavea: Latency-aware
  video analytics on edge computing platform,'' in \emph{ACM/IEEE Symposium on
  Edge Computing}, 2017, pp. 1--13.

\bibitem{zhang2022blockchain}
M.~Zhang, J.~Cao, Y.~Sahni, Q.~Chen, S.~Jiang, and L.~Yang, ``Blockchain-based
  collaborative edge intelligence for trustworthy and real-time video
  surveillance,'' \emph{IEEE Transactions on Industrial Informatics}, 2022.

\bibitem{boyd2004convex}
S.~Boyd, S.~P. Boyd, and L.~Vandenberghe, \emph{Convex Optimization}.\hskip 1em
  plus 0.5em minus 0.4em\relax Cambridge University Press, 2004.

\bibitem{zeng2020distream}
X.~Zeng, B.~Fang, H.~Shen, and M.~Zhang, ``Distream: scaling live video
  analytics with workload-adaptive distributed edge intelligence,'' in
  \emph{ACM Conference on Embedded Networked Sensor Systems}, 2020, pp.
  409--421.

\bibitem{jacob2018quantization}
B.~Jacob, S.~Kligys, B.~Chen, M.~Zhu, M.~Tang, A.~Howard, H.~Adam, and
  D.~Kalenichenko, ``Quantization and training of neural networks for efficient
  integer-arithmetic-only inference,'' in \emph{IEEE Conference on Computer
  Vision and Pattern Recognition}, 2018, pp. 2704--2713.

\bibitem{yang2021tree}
L.~Yang, Y.~Lu, J.~Cao, J.~Huang, and M.~Zhang, ``E-tree learning: A novel
  decentralized model learning framework for edge ai,'' \emph{IEEE Internet of
  Things Journal}, vol.~8, no.~14, pp. 11\,290--11\,304, 2021.

\bibitem{sandler2018mobilenetv2}
M.~Sandler, A.~Howard, M.~Zhu, A.~Zhmoginov, and L.-C. Chen, ``Mobilenetv2:
  Inverted residuals and linear bottlenecks,'' in \emph{IEEE Conference on
  Computer Vision and Pattern Recognition}, 2018, pp. 4510--4520.

\bibitem{hermans2017defense}
A.~Hermans, L.~Beyer, and B.~Leibe, ``In defense of the triplet loss for person
  re-identification,'' \emph{arXiv preprint arXiv:1703.07737}, 2017.

\bibitem{krajsa2011rtt}
O.~Krajsa and L.~Fojtova, ``Rtt measurement and its dependence on the real
  geographical distance,'' in \emph{2011 34th International Conference on
  Telecommunications and Signal Processing (TSP)}.\hskip 1em plus 0.5em minus
  0.4em\relax IEEE, 2011, pp. 231--234.

\bibitem{kjorveziroski2022kubernetes}
V.~Kjorveziroski and S.~Filiposka, ``Kubernetes distributions for the edge:
  serverless performance evaluation,'' \emph{The Journal of Supercomputing},
  pp. 1--28, 2022.

\bibitem{xiong2018extend}
Y.~Xiong, Y.~Sun, L.~Xing, and Y.~Huang, ``Extend cloud to edge with
  kubeedge,'' in \emph{IEEE/ACM Symposium on Edge Computing}, 2018, pp.
  373--377.

\bibitem{santos2019towards}
J.~Santos, T.~Wauters, B.~Volckaert, and F.~De~Turck, ``Towards network-aware
  resource provisioning in kubernetes for fog computing applications,'' in
  \emph{IEEE Conference on Network Softwarization}, 2019, pp. 351--359.

\bibitem{larsson2020impact}
L.~Larsson, W.~T{\"a}rneberg, C.~Klein, E.~Elmroth, and M.~Kihl, ``Impact of
  etcd deployment on kubernetes, istio, and application performance,''
  \emph{Software: Practice and Experience}, vol.~50, no.~10, pp. 1986--2007,
  2020.

\bibitem{han2019ondisc}
Z.~Han, H.~Tan, X.-Y. Li, S.~H.-C. Jiang, Y.~Li, and F.~C. Lau, ``Ondisc:
  Online latency-sensitive job dispatching and scheduling in heterogeneous
  edge-clouds,'' \emph{IEEE/ACM Transactions on Networking}, vol.~27, no.~6,
  pp. 2472--2485, 2019.

\bibitem{sundar2018offloading}
S.~Sundar and B.~Liang, ``Offloading dependent tasks with communication delay
  and deadline constraint,'' in \emph{IEEE Conference on Computer
  Communications}, 2018, pp. 37--45.

\bibitem{wang2021dependent}
J.~Wang, J.~Hu, G.~Min, W.~Zhan, A.~Zomaya, and N.~Georgalas, ``Dependent task
  offloading for edge computing based on deep reinforcement learning,''
  \emph{IEEE Transactions on Computers}, 2021.

\bibitem{sahni2018data}
Y.~Sahni, J.~Cao, and L.~Yang, ``Data-aware task allocation for achieving low
  latency in collaborative edge computing,'' \emph{IEEE Internet of Things
  Journal}, vol.~6, no.~2, pp. 3512--3524, 2018.

\bibitem{sahni2020multi}
Y.~Sahni, J.~Cao, L.~Yang, and Y.~Ji, ``Multi-hop multi-task partial
  computation offloading in collaborative edge computing,'' \emph{IEEE
  Transactions on Parallel and Distributed Systems}, vol.~32, no.~5, pp.
  1133--1145, 2020.

\bibitem{baktir2017can}
A.~C. Baktir, A.~Ozgovde, and C.~Ersoy, ``How can edge computing benefit from
  software-defined networking: A survey, use cases, and future directions,''
  \emph{IEEE Communications Surveys \& Tutorials}, vol.~19, no.~4, pp.
  2359--2391, 2017.

\bibitem{li2018adaptive}
X.~Li, D.~Li, J.~Wan, C.~Liu, and M.~Imran, ``Adaptive transmission
  optimization in sdn-based industrial internet of things with edge
  computing,'' \emph{IEEE Internet of Things Journal}, vol.~5, no.~3, pp.
  1351--1360, 2018.

\bibitem{wang2019software}
A.~Wang, Z.~Zha, Y.~Guo, and S.~Chen, ``Software-defined networking enhanced
  edge computing: A network-centric survey,'' \emph{Proceedings of the IEEE},
  vol. 107, no.~8, pp. 1500--1519, 2019.

\end{thebibliography}

\end{document}